\documentclass[11pt,twoside]{article} 
\usepackage{times,fancyhdr}
\usepackage{cite}
\usepackage{graphicx}

\date{}

\title{Magnetoexciton dispersion in GaAs-(Ga,Al)As single and coupled quantum wells} 
\author{Z. G. Koinov} 

\begin{document} 

\pagestyle{fancy}
\fancyhead{} 
\fancyhead[EC]{} \fancyhead[EL,OR]{\thepage} \fancyhead[OC]{}
\fancyfoot{} 
\renewcommand\headrulewidth{0.5pt}
\addtolength{\headheight}{2pt} 

\maketitle 
          We discuss magnetoexcitons dispersion in single and coupled
   $GaAs-(Ga,Al)As$ quantum wells using the Bethe-Salpeter (B-S) formalism.
   The B-S formalism in the case of quantum wells provides an equation for the exciton wave
   function which depends on two space variables plus the time variable, i.e. the B-S equation is $2+1$-dimensional equation.
       We compare the results for magnetoexcitons dispersion, obtained in the LLL approximations with
   the results calculated by solving the exact B-S equation. It is shown that the exact B-S equation has an extra term
    (B-S term) that does not exist in the LLL approximation. Within the framework of the variational method, we obtain that,
     (i) the ground-state energy of a heavy-hole magnetoexciton with a zero wave vector in $GaAs-(Ga,Al)As$ quantum wells,
     calculated by means of the exact B-S equation, is very close to the ground-state energy, obtained in the LLL approximation,
     (ii) in a strong perpendicular magnetic field the magnetoexciton dispersion (in-plane  magnetoexciton mass) is
     determined mainly by the B-S term rather than the term that describes the electron-hole Coulomb interaction in the LLL approximation.

\section{Schr\"{o}dinger equation for magnetoexcitons in quantum wells}
The bound states between two charged fermions, an electron from
the conductive band and a hole from the valence band, in the
presence of a magnetic field are called magnetoexcitons. In what
follows we consider a single quantum well (SQW)  and coupled
quantum wells (CQW's) made with direct-gap semiconductor that has
nondegenerate and isotropic bands:
${E}_c(\textbf{k},k_z)=\textit{E}_g
+\hbar^2\textbf{k}^2/2m_c+\hbar^2k_z^2/2m_{c}$ and
${E}_v(\textbf{k},k_z)=\hbar^2\textbf{k}^2/2m_v+\hbar^2k_z^2/2m_{v}$,
where \textbf{k} is a two-dimensional (2D) wave vector,
$\textit{E}_g$ is the semiconductor band gap, and $m_c$ ($m_v$) is
the electron (hole) effective mass.  The z-axis is chosen to be
the axis of growth of the quantum-well structure, and the constant
magnetic fields is $\textbf{B}=(0,0,B)$. The x-y plane has been
taken to be the plane of confinement. In what follows we neglect
any electron-hole correlations along the z-axis. This
approximation takes place when the effective mass of the hole
considerably exceeds that of the electron and the slow motion of
the hole is separated from the fast motion of the electron. The
assumption is applicable for many crystals of $A^{III}B^{V}$ type.
In the presence of confinement potentials $U_{c,v}(z)$, the
corresponding electron $\varphi$ and hole $\phi$ wave functions
are defined by the solutions of the one-particle Schr\"{o}dinger
equations:
$$-\frac{\hbar^2}{2m_{c}}\frac{d^{2}\varphi_{\lambda}}{dz_c^{2}}+
U_c(z_c)\varphi_\lambda(z_c)=E_{\lambda
c}\varphi_\lambda(z_c),$$$$
-\frac{\hbar^2}{2m_{v}}\frac{d^{2}\phi_{\xi}}{dz_v^{2}}+U_v(z_v)\phi_\xi(z_v)=E_{\xi
v} \phi_\xi(z_v).$$
 Here, $E_{\lambda c} (E_{\xi v})$ is the electron (or hole)
confinement energy, $\lambda$ and $\xi$ denote the quantum numbers
of the states in the confinement potential. For simplicity, we
shall take into account only the first electron $E_{0c}$ and hole
$E_{0v}$ confinement levels. In the above equation $z_c$ and $z_v$
are the electron and hole z-coordinates, respectively. \\
The exciton motion in $(x,y)$-plane changes its spectrum, i.e. the
 magnetoexciton energy $E(\textbf{Q})$ depends on the in-plane exciton pseudomomentum  $\hbar\textbf{Q}=\hbar
 (Q_x,Q_y,0)$.
The influence of the exciton motion on its spectrum (the
dispersion relation), in the case
    of a SQW or CQW's have been studied
extensively over the past decades \cite{ Shevchenko97,
Lozovik97,Dzyubenko,Shevchenko98, Lozovik99}, assuming that the
 magnetoexciton energy $E(\textbf{Q})$ and the corresponding
 wave functions of the relative internal
motion can be obtained from the solutions of the Schr\"{o}dinger
 equation with the following Hamiltonian:
 \begin{equation}
\widehat{H}=E_g+E_{0c}+E_{0v}-\frac{\hbar^2}{2\mu}\nabla^2_\textbf{r}+\frac{\imath
e \gamma \hbar}{2\mu c}(\textbf{B}\times
\textbf{r})\textbf{.}\nabla_\textbf{r}+\frac{e^2B^2}{8\mu
c^2}\textbf{r}^2- V_C(\textbf{r}+\textbf{R}_0). \label{Hperp}
\end{equation}
Here, $\textbf{r}=\textbf{r}_c-\textbf{r}_v$, and
$\textbf{r}_{c(v)}$ is the two-dimensional (2D) electron (hole)
position vector. $\mu=m_cm_v/M$ is the exciton reduced mass, where
$M=m_c+m_v$ is the exciton in-plane mass. $\gamma=(m_v-m_c)/M$,
  $\textbf{R}_0=l^2\textbf{Q}_0$, where
  $\textbf{Q}_0=(-Q_y,Q_x,0)$,
and $l = (\hbar c/e B)^{1/2}$ is the magnetic length. $V_C$
represents the electron-hole Coulomb attraction screened by the
high-frequency dielectric constant $\epsilon_\infty$:
\begin{equation}V_C(\textbf{r})=\frac{2\pi e^2 }{\epsilon_\infty
}\int\frac{d^{2}\textbf{q}}{(2 \pi)^{2}}\frac{
f(|\textbf{q}|)}{|\textbf{q}|}\exp\left(\imath\textbf{q.r}\right),\label{Pot0}\end{equation}
where the structure factor  $f$ is defined by:
\begin{equation}
f(|\textbf{q}|)=f(q)=\int_{-\infty}^{+\infty}dz_c\int_{-\infty}^{+\infty}dz_v
\exp\{-q(z_c-z_v)]\}\varphi_{0c}^2(z_c)\phi_{0v}^2(z_v)
.\label{SF0}\end{equation}
 Since the last
term in (\ref{Hperp}) is the only term which depends on the
exciton momentum, the following statements take place: \\ (i) The
 magnetoexciton dispersion does not depend on the electron and hole masses. \\ (ii) The
 magnetoexciton dispersion is determined only by Coulomb interaction.\\

\section{Dimensional reduction in the dynamics of bulk magnetoexcitons}

Strictly speaking, the excitons are bound states between two
charged fermions, and therefore, the appropriate framework for the
description of the bound states is the Bethe-Salpeter (B-S)
formalism\cite{BS,GL,W, C}. In the absence of a magnetic field, by
using a series of approximations (such as the introduction of the
equal-time wave function, the assumption that the B-S kernel
depends only on the difference of the relative momenta) the B-S
equation for electron-hole bound states can be simplified to the
well-known Schr\"{o}dinger equation for the relative internal
motion\cite{P}. The existence of a magnetic field induces a
coupling between the center-of-mass and the relative internal
motions, because even a small transverse exciton
 velocity (or small transverse wave vector \textbf{Q})
will induce an electric field in the rest frame of the exciton
which will push the electron and the hole apart, so the binding
energy must decrease as the transverse velocity increases. Thus,
one can expect that in the presence of a magnetic field the
simplification of the B-S equation to the Schr\"{o}dingier
equation is not trivial.\\ Several non-trivial effects produced by
magnetic fields have been recently predicted in quantum field
theories. For example, in the massless QED, by means of the lowest
Landau level (LLL) approximation, the B-S equation has been
reduced to the Schr\"{o}dingier equation, and as a result, it was
predicted that the external constant magnetic field generates an
energy gap (dynamical mass) in the spectrum of massless fermions
for any arbitrary weak attractive interaction between
fermions\cite{MC,MC1,L,H}. This effect is model independent
(universal), because the physical reason of this effect lies in
the fact that dynamics of the LLL is essentially
$D-2$-dimensional. In other words, the essence of this effect is
the dimensional reduction (from $3+1$ to $1+1$, or $2+1\rightarrow
0+1$) in the dynamics of fermion pairing in the presence of a the
constant magnetic field. Later, it was suggested that a similar
effect could explain some experimental findings in the physics of
high-temperature layered superconductors\cite{FM}. In what
follows, we will see that the dimensional reduction in the
dynamics of magnetoexcitons manifests itself in the fact that the
magnetoexciton dispersion does not depend on the electron and hole
masses.\\ We first use the B-S formalism to describe excitons in a
bulk material in the presence of a strong constant magnetic field
$\textbf{B}$ along the z-axis. After that, we apply the bulk B-S
formalism to a SQW or CQW's. The process of generalizing the bulk
equations to the case of quantum-well structures is a
straightforward procedure because of the assumption that there are
no electron-hole correlations along the z-axis.
\\The basic assumption in
the B-S formalism is that the electron-hole bound states are
described by the B-S wave function (B-S amplitude)
$\Psi(1;2)=\Psi(\textbf{r}_c,\textbf{r}_v;z_c,z_v;t_1,t_2)$, where
the variables $1$ and $2$ represent the corresponding coordinates
and the time variables. This function determines the probability
amplitude to find the electron at the point $(\textbf{r}_c,z_c)$
at the moment $t_1$ and the hole at the point $(\textbf{r}_v,z_v)$
at the moment $t_2$. The B-S amplitude satisfies the following
equation:
\begin{equation}\Psi(1;2)=\int d(1',2',1",2") G_c(1;1')G_v(2';2)I\left(%
\begin{array}{cc}
   1' & 1"  \\
  2' & 2"\\
\end{array}%
\right)\Psi(1";2").\label{BSz1}\end{equation} Here $I$ is the
irreducible B-S kernel, and $G_{c,v}$ are the electron and the
hole Green's functions. If the screening effects are taken into
account by means of the high-frequency dielectric constant
$\epsilon_\infty$, then the irreducible kernel is given by
\begin{equation}V(\textbf{r};z)=-\frac{4\pi e^2}{\epsilon_\infty}\int\frac{d^2\textbf{q}}{(2\pi)^2}\frac{dq_z}{2\pi}
\frac{1}{|\textbf{q}|^2+q_z^2}
\exp\left[\imath\left(\textbf{q.r}+q_zz\right)\right].\label{Pot}\end{equation}
In what follows, we use the center-of-mass
$(\textbf{R},Z)=(\alpha_c\textbf{r}_c+\alpha_v\textbf{r}_v,\alpha_cz_c+\alpha_vz_v)$
and the relative
$(\textbf{r},z)=(\textbf{r}_c-\textbf{r}_v,z_c-z_v)$ coordinates.
 The coefficients $\alpha_c=(1-\gamma)/2$, $\alpha_v=(1+\gamma)/2$ are
expressed in terms of the parameter $\gamma=(m_v-m_c)/(m_c+m_v)$
which accounts for the difference between the electron and the
hole masses. The B-S equation for the equal-time B-S amplitude in
the center-of-mass and reduced coordinates assumes the form:
\begin{eqnarray}
&\Psi_{\textbf{Q},Q_z}(\textbf{r},\textbf{R};z,Z;t,t)=\int dz'dZ'
d^2\textbf{r}'d^2\textbf{R}'dt_1dt_2\nonumber\\&G_c(\textbf{R}+\alpha_v\textbf{r},\textbf{R}'+\alpha_v\textbf{r}';Z+\frac{m_{vz}}{M_z}z,Z'+\frac{m_{vz}}{M_z}z';t-t_1)
\nonumber\\&G_v(\textbf{R}'-\alpha_c\textbf{r}',\textbf{R}-\alpha_c\textbf{r};Z'-\frac{m_{cz}}{M_z}z',Z-\frac{m_{cz}}{M_z}z;t_1-t)\nonumber\\&
V(\textbf{r}';z')\Psi_{\textbf{Q},Q_z}(\textbf{r}',\textbf{R}';z',Z';t_1,t_1).
\label{BSEq}\end{eqnarray}  The B-S amplitude depends on the
relative internal time $t-t'$ and on the "center-of-mass" time:
\begin{equation}\Psi_{\textbf{Q},Q_z}(\textbf{r},\textbf{R};z,Z;t,t')=\exp\left(-\frac{\imath
E(\textbf{Q},Q_z)}{\hbar}
(\alpha_ct+\alpha_vt')\right)\psi_{\textbf{Q},Q_z}(\textbf{r},\textbf{R};z,Z;t-t'),\label{Ps}\end{equation}
where $E(\textbf{Q},Q_z)$ is the exciton dispersion. Introducing
the time Fourier-transforms according to the rule
$f(t)=\int_{-\infty}^\infty f(\omega)\exp\left(\imath\omega
t\right)\frac{d\omega}{2\pi}$, we transform the B-S equation into
the following form:
\begin{eqnarray}
&\psi_{\textbf{Q},Q_z}(\textbf{r},\textbf{R};z,Z;\omega)=\int
dz'dZ'
d^2\textbf{r}'d^2\textbf{R}'\frac{d\Omega}{2\pi}\nonumber\\&G_c\left(\textbf{R}+\alpha_v\textbf{r},\textbf{R}'+
\alpha_v\textbf{r}';Z+\alpha_vz,Z'+\alpha_vz';\hbar\omega+\alpha_cE(\textbf{Q},Q_z)\right)
\nonumber\\&G_v\left(\textbf{R}'-\alpha_c\textbf{r}',\textbf{R}-\alpha_c\textbf{r};Z'-\alpha_cz',Z-\alpha_cz;\hbar\omega-\alpha_vE(\textbf{Q},Q_z)\right)\nonumber\\&
V(\textbf{r}';z')\psi_{\textbf{Q},Q_z}(\textbf{r}',\textbf{R}';z',Z';\Omega).
\label{BSEq1}\end{eqnarray}
 where
$\psi_{\textbf{Q},Q_z}(\textbf{r},\textbf{R};z,Z;\Omega)$ is the
Fourier transform of
$\psi_{\textbf{Q},Q_z}(\textbf{r},\textbf{R};z,Z;t)$. Since the
translation symmetry is broken by the magnetic field, the Green's
functions can be written as a product of phase factors and
 translation invariant parts. The phase factor depends on the
gauge. In the symmetric gauge the vector potential of the magnetic
field $\textbf{A}$ is defined by
$\textbf{A}(\textbf{r})=(1/2)\textbf{B}\times\textbf{r}$, and the
Green's functions are\cite{DL}:
\begin{equation}
G_{c,v}(\textbf{r},\textbf{r}';z,z';\omega)=\exp\left[\imath\frac{e}{\hbar
c}\textbf{r.A}(\textbf{r}')\right]\widetilde{G}_{c,v}(\textbf{r}-\textbf{r}';z-z';\omega).
\label{GF}\end{equation}
\\The broken  translation symmetry requires a
phase factor for the B-S amplitude:
\begin{equation}\psi_{\textbf{Q},Q_z}(\textbf{r},\textbf{R};z,Z;\Omega)=\exp\left[\imath\frac{e}{\hbar
c}\textbf{r.A}(\textbf{R})\right]\chi_{\textbf{Q},Q_z}(\textbf{r},\textbf{R};z,Z;\Omega).\label{BSA}\end{equation}
 The B-S equation (\ref{BSEq1}) admits translation invariant
 solution of the form:
\begin{equation}
\chi_{\textbf{Q},Q_z}(\textbf{r},\textbf{R};z,Z;\omega)=\exp\left[-\imath\left(
\textbf{Q.R}+Q_zZ\right)\right]
\widetilde{\chi}_{\textbf{Q},Q_z}(\textbf{r};z;\omega).\label{BSF1}\end{equation}
The function
$\widetilde{\chi}_{\textbf{Q},Q_z}(\textbf{r};z;\omega)$ satisfies
the following B-S equation:
\begin{eqnarray}
&\widetilde{\chi}_{\textbf{Q},Q_z}(\textbf{r};z;\omega)=\int
dz'dZ'
d^2\textbf{r}'d^2\textbf{R}'\frac{d\Omega}{2\pi}\exp\left[\frac{\imath
e}{\hbar
c}\left((\textbf{r}+\textbf{r}')\textbf{.A}(\textbf{R}'-\textbf{R})+\gamma\textbf{r.A}(\textbf{r}')\right)\right]\nonumber\\
&\widetilde{G}_c(\textbf{R}-\textbf{R}'+\alpha_v(\textbf{r}-\textbf{r}');Z-Z'+\alpha_v(z-z');\hbar\omega+\alpha_cE)
\nonumber\\&\widetilde{G}_v(\textbf{R}'-\textbf{R}+\alpha_c(\textbf{r}-\textbf{r}');Z'-Z+\alpha_c(z-z');\hbar\omega-\alpha_vE)
\nonumber\\&V(\textbf{r}';z')\widetilde{\chi}_{\textbf{Q},Q_z}(\textbf{r}';z';\Omega).
\label{BSEq2}\end{eqnarray} The substitution
$\textbf{R}'\rightarrow \textbf{R}'+\textbf{R}+\gamma \textbf{r}$
provides the following equation for the Fourier transform of the
exciton wave function
 $\widetilde{\chi}_{\textbf{Q},Q_z}(\textbf{k};k_z;\omega)=\int dz
d^2\textbf{r}\exp-\imath\left(\textbf{k.r}+k_zz\right)\widetilde{\chi}_{\textbf{Q},Q_z}(\textbf{r};z;\omega)$
of the exciton wave function:
\begin{eqnarray}
&\widetilde{\chi}_{\textbf{Q},Q_z}(\textbf{k}-\frac{\gamma}{2}\textbf{Q};k_z;\omega)=\int
\frac{dp_z}{2\pi}
\frac{d^2\textbf{q}}{(2\pi)^2}\frac{d^2\textbf{p}}{(2\pi)^2}d^2\textbf{R}\int_{-\infty}^\infty\frac{d\Omega}{2\pi}\exp\left[-\imath
(\textbf{q}+\textbf{Q})\textbf{.R}\right]\times\nonumber\\
&\widetilde{G}_c\left(\frac{1}{2}\textbf{q}+\textbf{k}
-\frac{e}{\hbar
c}\textbf{A}(\textbf{R});k_z+\alpha_vQ_z;\hbar\omega+\alpha_cE\right)
\times\nonumber\\&\widetilde{G}_v\left(-\frac{1}{2}\textbf{q}+\textbf{k}
-\frac{e}{\hbar
c}\textbf{A}(\textbf{R});k_z-\alpha_cQ_z;\hbar\omega-\alpha_vE\right)\times\nonumber\\&
V\left(\textbf{p}-\left[\textbf{k} -\frac{2e}{\hbar
c}\textbf{A}(\textbf{R})\right];p_z-k_z\right)
\widetilde{\chi}_{\textbf{Q},Q_z}(\textbf{p}-\frac{\gamma}{2}\textbf{Q};p_z;\Omega),
\label{BSEqk1}\end{eqnarray} where $V(\textbf{k};k_z)=-\left(4\pi
e^2/\varepsilon_\infty\right)\left(\textbf{k}^2+k_z^2\right)^{-1}$
and $\widetilde{G}_{c,v}\left(\textbf{k};k_z ;\hbar\omega\right)$
are the Fourier transforms of
$\widetilde{G}_{c,v}\left(\textbf{r};z ;\hbar\omega\right)$. \\
In the effective-mass approximation the exact fermion Green's
functions $G_{c,v}$ are replaced by the corresponding propagator
of the free fermions $G^{(0)}_{c,v}$. The translation invariant
parts $\widetilde{G}^{(0)}_{c,v}$ can be decomposed over the
Landau level poles:
\begin{eqnarray}
&\widetilde{G}^{(0)}_{c,v}(\textbf{r};z;\hbar\omega)=\int
\frac{d^2\textbf{k}}{(2\pi)^2}\frac{dk_z}{2\pi}\widetilde{G}^{(0)}_{c,v}(\textbf{k};k_z;\hbar\omega)
\exp\imath\left(\textbf{k.r}+k_zz\right),
\nonumber\\&\widetilde{G}^{(0)}_{c,v}(\textbf{k};k_z;\hbar\omega)=2\sum_{n=0}^\infty(-1)^n\exp\left(-l^2\textbf{k}^2\right)L_n\left(2l^2\textbf{k}^2\right)
\times\nonumber\\&\left(\hbar\omega-\left[\hbar^2
k_z^2/2m_{c}+E_g+\hbar\Omega_{c}(n+1/2)\right] +\imath
0^+\right)^{-1}\nonumber\\&
\widetilde{G}^{(0)}_{v}(\textbf{k};k_z;\hbar\omega)=2\sum_{n=0}^\infty(-1)^n\exp\left(-l^2\textbf{k}^2\right)L_n\left(2l^2\textbf{k}^2\right)\times\nonumber\\&
\left(\hbar\omega+\left[\hbar^2
k_z^2/2m_{v}+\hbar\Omega_{v}(n+1/2)\right] -\imath
0^+\right)^{-1}. \label{GF1}\end{eqnarray} Here $L_n(x)$ are the
Laguerre polynomials, and $\hbar\Omega_{c,v}=\hbar e B/cm_{c,v}$
are the electron and hole cyclotron energies. In strong magnetic
fields the probability for transitions to the excited Landau
levels due to the Coulomb interaction is small. Thus, the
contributions to the Green's functions from the excited Landau
levels is negligible, and therefore, one can apply the lowest
Landau level (LLL) approximation, where we keep only $n=0$ term in
(\ref{GF1}):
\begin{eqnarray}&
\widetilde{G}_{c}(\textbf{k};k_z;\hbar\omega)\approx
2\exp\left(-l^2\textbf{k}^2\right)
\left(\hbar\omega-\left[E_g+\hbar^2
k_z^2/2m_{c}+\hbar\Omega_{c}/2\right] +\imath
0^+\right)^{-1},\nonumber\\&
\widetilde{G}_{v}(\textbf{k};k_z;\hbar\omega)\approx
2\exp\left(-l^2\textbf{k}^2\right)\left(\hbar\omega+\left[\hbar^2
k_z^2/2m_{v}+\hbar\Omega_{v}/2\right] -\imath 0^+\right)^{-1}.
\label{LLL}\end{eqnarray}The solution of the B-S equation in the
LLL approximation can be written in the following form:
\begin{equation}
\widetilde{\chi}_{\textbf{Q},Q_z}(\textbf{k};k_z;\omega)=
\exp\left[-l^2\left(\textbf{k}+\frac{\gamma}{2}\textbf{Q}\right)^2-\imath\textbf{R}_0\textbf{.k}\right]\Phi_{Q_z}(k_z;\omega).
\label{Sol}\end{equation} Thus,  the LLL approximation reduces the
problem from $3+1$ dimensions to $1+1$ dimensions problem for
obtaining functions $\Phi_{Q_z}(k_z;\omega)$ and the energy
$E(\textbf{Q},Q_z)$ from the following equation:
\begin{eqnarray}&\Phi_{Q_z}(k_z;\omega)=\int
\frac{dp_z}{2\pi}\frac{d\Omega}{2\pi}I_\textbf{Q}(p_z-k_z)\Phi_{Q_z}(p_z;\Omega)\nonumber\\&
\left[\frac{1}{\hbar\omega+\alpha_cE-\left(E_g+\frac{\hbar^2}{2m_c}(k_z+\alpha_cQ_z)^2+\frac{\hbar\Omega_c}{2}\right)+\imath0^+}+
\frac{1}{\hbar\omega-\alpha_vE+\frac{\hbar^2}{2m_v}(k_z-\alpha_vQ_z)^2+\frac{\hbar\Omega_v}{2}-\imath0^+}\right].
\label{Sol1}\end{eqnarray} In the LLL approximation, the in-plane
exciton dispersion is determined by the Coulomb interaction:
\begin{equation}
I_\textbf{Q}(k_z)=\frac{4\pi e^2}{\varepsilon_\infty}\int
d^2\textbf{r}\frac{d^2\textbf{q}}{(2\pi)^2}\psi^2_{00}(\textbf{r})\frac{\exp\left[\imath\textbf{q.}(\textbf{r}+\textbf{R}_0)\right]}{(q^2+k_z^2)}
.\label{Disp}\end{equation} Here,
$\psi_{00}(\textbf{r})=\frac{1}{\sqrt{2\pi}l}\exp\left(-r^2/4l^2\right)$
is the ground-state wave function of a hydrogen atom in a magnetic
field. The solution of (\ref{Sol1}) can be chosen in the following
form:
\begin{eqnarray}
&\Phi_{Q_z}(k_z,\omega)=\phi_{Q_z}(k_z)\left[
\hbar\omega+\alpha_cE-\left[E_g+\frac{\hbar^2}{2m_c}(k_z+\alpha_cQ_z)^2+\frac{\hbar\Omega_c}{2}\right]+\imath0^+
\right]^{-1}\times\nonumber\\&\left[\hbar\omega-\alpha_vE+\left[\frac{\hbar^2}{2m_v}(k_z-\alpha_vQ_z)^2+\frac{\hbar\Omega_v}{2}\right]-\imath0^+
\right]^{-1}, \label{Sol2}\end{eqnarray} where $\phi_{Q_z}(k_z)$
is a function to be determined. By integrating both sides of
(\ref{Disp}) over $\omega$, we find the following equation for the
exciton wave function
$$\Phi_{Q_z}(k_z)=\int\frac{d\omega}{2\pi}\Phi_{Q_z}(k_z,\omega)=\phi_{Q_z}(k_z)/\left(E-E_g-\hbar^2k_z^2/2\mu-\hbar^2Q_z^2/2M\right)$$
and exciton energy
$E_b(\textbf{Q},Q_z)=E_g+\frac{1}{2}\hbar\Omega-E(\textbf{Q},Q_z)$
($\Omega=\hbar eB/\mu$ is the exciton cyclotron energy):
\begin{equation}
0=\left(\frac{\hbar^2k^2_z}{2\mu}+\frac{\hbar^2Q^2_z}{2M}+E_b(\textbf{Q},Q_z)\right)\Phi_{Q_z}(k_z)
-\int \frac{dp_z}{2\pi}I_\textbf{Q}(k_z-p_z)\Phi_{Q_z}(p_z)
.\label{ExcD}\end{equation} The exciton binding energy $E_b>0$
could be obtained from the solutions of (\ref{ExcD}) by means of
$E_b=E_b(\textbf{Q}=0,Q_z=0)$.\\
In the case when $\textbf{Q}=0$ and $Q_z=0$, eq. (\ref{ExcD})
  is similar to the well-known one-dimensional Schr\"{o}dinger
  equation for a hydrogen atom in the adiabatic approximation
  \cite{HA,HA1,HA2,HA3}.
\section{Magnetoexciton dispersion in quantum wells in the lowest Landau level approximation}
The assumptions that: (i) we neglect any electron-hole
correlations along the z-axis, and (ii) we take into account only
the first electron $E_{0c}$ and hole $E_{0v}$ confinement levels
with wave functions $\varphi_{0c}(z_c)$ and $\phi_{0v}(z_v)$,
respectively, greatly simplify the description of the motion along
the z-axis. In the cases of a SQW and CQW's, the Fourier transform
of the exciton wave function satisfies the following B-S equation:
\begin{eqnarray}
&\widetilde{\chi}_\textbf{Q}(\textbf{k}-\frac{\gamma}{2}\textbf{Q};\omega)=\int
\frac{d^2\textbf{q}}{(2\pi)^2}\frac{d^2\textbf{p}}{(2\pi)^2}d^2\textbf{R}\int_{-\infty}^\infty\frac{d\Omega}{2\pi}\exp\left[-\imath
(\textbf{q}+\textbf{Q})\textbf{.R}\right]\times\nonumber\\
&\widetilde{G}_c\left(\frac{1}{2}\textbf{q}+\textbf{k}
-\frac{e}{\hbar
c}\textbf{A}(\textbf{R});\hbar\omega+\alpha_cE\right)
\widetilde{G}_v\left(-\frac{1}{2}\textbf{q}+\textbf{k}
-\frac{e}{\hbar
c}\textbf{A}(\textbf{R});\hbar\omega-\alpha_vE\right)\times\nonumber\\&
V\left(\textbf{p}-\left[\textbf{k} -\frac{2e}{\hbar
c}\textbf{A}(\textbf{R})\right]\right)\widetilde{\chi}_\textbf{Q}(\textbf{p}-\frac{\gamma}{2}\textbf{Q};\Omega),
\label{2BSEqk1}\end{eqnarray} where the potential
$V(\textbf{k})=-\left(2\pi
e^2f(|\textbf{k}|)/\varepsilon_\infty\right)|\textbf{k}|^{-1} $
depends on the quantum-well geometry through the structure
factor$f(\textbf{k})$.\\In the LLL approximation the exact fermion
Green's functions $G_{c,v}$ are replaced by the corresponding
propagator of the free fermions $G^{(0)}_{c,v}$:
\begin{eqnarray}&
\widetilde{G}_{c}(\textbf{k};\hbar\omega)\approx
2\exp\left(-l^2\textbf{k}^2\right)\left(\hbar\omega-[E_g+E_{0c}+\hbar\Omega_{c}/2]+\imath
0^+\right)^{-1}\nonumber,\\&\widetilde{G}_{v}(\textbf{k};\hbar\omega)\approx
2\exp\left(-l^2\textbf{k}^2\right)\left(\hbar\omega+E_{0v}+\hbar\Omega_{v}/2
-\imath 0^+\right)^{-1}. \label{2LLL}\end{eqnarray}The solution of
the B-S equation in the LLL approximation can be written in the
following form:
\begin{equation}
\widetilde{\chi}_\textbf{Q}(\textbf{k};\omega)=\exp\left[-l^2\left(\textbf{k}+\frac{\gamma}{2}\textbf{Q}\right)^2-\imath\textbf{R}_0\textbf{.k}\right]\Phi_E(\omega).
\label{2Sol}\end{equation} Thus,  the LLL approximation reduces
the problem from $2+1$-dimensions to $0+1$-dimension problem. The
function $\Phi_E(\omega)$ energy $E(\textbf{Q})$ can be obtained
from the following B-S equation:
\begin{eqnarray}
&\Phi_E(\omega)=-I(|\textbf{Q}|)\int_{-\infty}^{\infty}\frac{d\Omega}{2\pi}\Phi_E(\Omega)
\nonumber\times\\&\left(\hbar\omega+\alpha_cE-E_g-E_{0c}-\hbar\Omega_c/2+\imath0^+\right)^{-1}
\left(\hbar\omega-\alpha_vE+E_{0v}+\hbar\Omega_v/2-\imath0^+\right)^{-1}
 \label{Z}\end{eqnarray} In the LLL approximation, the
exciton dispersion is determined by the term:
\begin{equation}I(\textbf{Q})=\frac{2\pi e^2}{\varepsilon_\infty}\int
d^2\textbf{r}\frac{d^2\textbf{q}}{(2\pi)^2}\psi^2_{00}(\textbf{r})\frac{f(|\textbf{q}|)\exp\left[\imath\textbf{q.}(\textbf{r}+\textbf{R}_0)\right]}{|\textbf{q}|}
.\label{2Disp}
\end{equation}
The solution $\Phi_E(\omega)$ of (\ref{Z}) can be chosen in the
following form:
\begin{eqnarray}
&\Phi_E(\omega)=\nonumber\\&\left[\left(\hbar\omega+\alpha_cE-E_g-E_{0c}-\frac{\hbar\Omega_c}{2}+\imath0^+\right)
\left(\hbar\omega-\alpha_vE+E_{0v}+\frac{\hbar\Omega_v}{2}-\imath0^+\right)\right]^{-1}.
\label{2Sol2}\end{eqnarray} Integrating both sides of B-S equation
(\ref{Z}) over $\omega$, we find that the exciton dispersion is
determined only by the Coulomb interaction (\ref{2Disp}):
\begin{equation}E(|\textbf{Q}|)=E_g+E_{0c}+E_{0v}+\hbar\Omega/2-I(|\textbf{Q}|).
\label{2ExcD}\end{equation} It turns out that in the LLL
approximation the magnetoexciton dispersion does not depend on the
electron and hole masses and is determined only by Coulomb
interaction.\\ The LLL approximation greatly simplifies the
equations, but we may ask whether the magnetoexciton dispersion
will be significantly affected by the contributions from the
\emph{infinity number} of Landau levels with indexes $n\geq 1$
neglected in the LLL approximation. In the next Section we address
this question.
\section{Magnetoexciton dispersion in $GaAs-(Ga,Al)As$ quantum wells}
In the previous two Sections, we decomposed the single-particle
electron (hole) Green's function over the Landau poles and we kept
only the term with index $n=0$. This term is relatively simple,
and allows us to perform all integrations in the B-S equation
(\ref{BSEqk1}). Unfortunately, the terms with $n\geq 1$ are more
complicated, and it is impossible to perform the integrations over
the corresponding variables. \\
There exists another approach which allows us to figure out the
contributions to magnetoexciton dispersion due to the Landau
levels with indexes $n\geq 1$.  It starts from the B-S equation
(\ref{BSz1}), but rewritten in the following form\cite{Z1,Z2}:
\begin{eqnarray}
&\left(\imath\hbar\frac{\partial}{\partial t_1} - \textit{E}_g -
\frac{1}{2 m_c} \left[-\imath \hbar\nabla _{\textbf{r}_c} +
\frac{e}{c} \textbf{A}(x_c,y_c,z _c ) \right]^2
-\frac{\hbar^2}{2m_{c}}\frac{\partial^{2}}{\partial z_c^{2}}
-U_c(z_c)\right)\times\nonumber\\&\left(\imath\hbar\frac{\partial}{\partial
t_2} - \frac{1}{2 m_v} \left[-\imath \hbar\nabla_{\textbf{r}_v} -
\frac{e}{c} \textbf{A} (x_v,y_v,z_v)\right]^2
-\frac{\hbar^2}{2m_{v}}\frac{\partial^{2}}{\partial
z_v^{2}}-U_v(z_v)\right)
\Psi(\textbf{r}_c,\textbf{r}_v;z_c,z_v;t_1,t_2)\nonumber \\&=
\imath V(\textbf{r} _c
-\textbf{r}_v;z_c-z_v)\Psi(\textbf{r}_c,\textbf{r}_v;z_c,z_v;t_1,t_1),\nonumber
\end{eqnarray}
where $V(\textbf{r},z)$ is defined by (\ref{Pot}). Since there are
no electron-hole correlations along the z-axis, we separate the
variables and write the B-S amplitude in the following form:
\begin{eqnarray}
&\Psi(\textbf{r}_c,z_c,t_1;\textbf{r}_v,z_v,t_2)=
\exp\left\{\imath\left[\textbf{Q.R}-\frac{e}{c\hbar}\textbf{r.A}(\textbf{R})-\frac{E}{\hbar}(\alpha_ct_1+\alpha_vt_2)\right]\right\}\times\nonumber\\&
\widetilde{\chi}_\textbf{Q}(\textbf{r};t_1-t_2)\varphi_0(z_c)\phi_0(z_v)
,\label{WF2}
\end{eqnarray}
where $E\equiv E(\textbf{Q})$ is the magnetoexciton dispersion.
After some tedious, but straightforward calculations, we arrive at
the conclusion that the  Fourier transform of the B-S amplitude
\begin{equation}
\widetilde{\chi}_\textbf{Q}(\textbf{r};t_1-t_2)=
\int\frac{d^2\textbf{q}}{(2\pi)^2}\int_{-\infty}^{+\infty}\frac{d\Omega}{2\pi}
\exp\left\{\imath[\textbf{q.r}-\Omega(t_1-t_2)]\right\}
\widetilde{\chi}_\textbf{Q}(\textbf{q};\Omega)\quad. \nonumber
\end{equation}
 satisfies the following equation\cite{Z1,Z2}:
\begin{eqnarray}
&\int \frac{d^2\textbf{q}'}{(2\pi)^2}\int
d^2\textbf{r}\exp\left(\imath(\textbf{q}'-\textbf{q})\textbf{.r}\right)
\left[\hbar\Omega-\Omega_c(\textbf{q}',\textbf{Q})-\Omega_c^{\textbf{B}_\perp}(\textbf{Q},\textbf{q}';\textbf{r})\right]\times\nonumber\\&
\left[\hbar\Omega-\Omega_v(\textbf{q}',\textbf{Q})-\Omega_v^{\textbf{B}_\perp}(\textbf{Q},\textbf{q}';\textbf{r})\right]
\widetilde{\chi}_{\textbf{Q}}(\textbf{q}';\Omega)\nonumber\\
&=-\imath \int\frac{d^{2}\textbf{q}'}{(2 \pi)^{2}}\frac{2\pi e^2
f(|\textbf{q}-\textbf{q}'|)}{\epsilon_\infty
|\textbf{q}-\textbf{q}'|}\int_{-\infty}^{+\infty} \frac{d
\Omega'}{2 \pi}\widetilde{\chi}_{\textbf{Q}}(\textbf{q}';\Omega')
. \label{BSE Perp}
\end{eqnarray}
Here, we use the following notations:
\begin{equation}
\Omega_c(\textbf{q},\textbf{Q})=E_c(\textbf{q}+\alpha_c\textbf{Q})+
E_{0c}-\alpha_cE,\quad
\Omega_v(\textbf{q},\textbf{Q})=-E_v(\textbf{q}-\alpha_v\textbf{Q})
-E_{0v}+\alpha_vE ,\label{Omega a}
\end{equation}
\begin{equation}
\Omega_c^{\textbf{B}_\perp}(\textbf{Q},\textbf{q};\textbf{r})=\frac{e\hbar}{2Mc}
(\textbf{B}_\perp\times\textbf{r})\textbf{.Q} +\frac{e\hbar }{2m_c
c}(\textbf{B}_\perp\times\textbf{r})\textbf{.} \textbf{q}+
\frac{e^2B_\perp^2}{8 m_c c^2}\textbf{r}^2 , \label{Omega b}
\end{equation}
\begin{equation}
\Omega_v^{\textbf{B}_\perp}(\textbf{Q},\textbf{q};\textbf{r})=\frac{e\hbar}{2Mc}
(\textbf{B}_\perp\times\textbf{r})\textbf{.Q} -\frac{e \hbar}{2m_v
c}(\textbf{B}_\perp\times\textbf{r})\textbf{.} \textbf{q}+
\frac{e^2B_\perp^2}{8 m_v c^2}\textbf{r}^2 , \label{Omega c}
\end{equation}
where  $E_{c,v}(\textbf{q})=E_{c,v}(\textbf{q},q_z=0)$. We are
looking for the solution of Eq. (\ref{BSE Perp}) of the form:
\begin{equation}
\widetilde{\chi}_{\textbf{Q}}(\textbf{q};\Omega)=\frac{g_{\textbf{Q}}(\textbf{q})}{[\hbar\Omega-\Omega_c(\textbf{q},\textbf{Q})+\imath
0^{+}][\hbar\Omega-\Omega_v(\textbf{q},\textbf{Q})-\imath 0^{+}]},
\label{W}
\end{equation}
where $g_\textbf{Q}(\textbf{q})$ is a function to be determined.\\
We introduce the function
$\widetilde{\chi}_{\textbf{Q}}(\textbf{q})$, which is the Fourier
transform of the equal-time B-S amplitude (or exciton wave
function)
$\widetilde{\chi}_{\textbf{Q}}(\textbf{r})=\widetilde{\chi}_{\textbf{Q}}(\textbf{r};t_1-t_2=0)$:
\begin{equation}
\widetilde{\chi}_{\textbf{Q}}(\textbf{q})=\int_{-\infty}^{+\infty}
\frac{d\Omega}{2\pi}\widetilde{\chi}_{\textbf{Q}}(\textbf{q};\Omega)\quad.\label{ExcWF}
\end{equation}
By taking into account the analytic properties of
$\widetilde{\chi}_{\textbf{Q}}(\textbf{q};\omega)$, we obtain the
following B-S equation for determining the exciton energy
$E'=E(\textbf{Q})-E_g-E_{0c}-E_{0v}$ and the Fourier transform of
the exciton wave function
$\widetilde{\chi}_{\textbf{Q}}(\textbf{q})$:
\begin{eqnarray}
&\int\frac{d^2\textbf{q}'}{(2\pi)^2}
\left[\left(\frac{\hbar^2\textbf{Q}^2}{2M}+\frac{\hbar^2\textbf{q}^2}{2\mu}\right)\delta(\textbf{q}-\textbf{q}')+
\Omega_c^{\textbf{B}}(\textbf{Q},\textbf{q},\textbf{q}')+\Omega_v^{\textbf{B}}(\textbf{Q},\textbf{q},\textbf{q}')
-\frac{2\pi e^2}{\epsilon_{\infty}}
\frac{f(|\textbf{q}-\textbf{q}'|)}{|\textbf{q}-\textbf{q}'|}\right]\times\nonumber\\&\widetilde{\chi}_{\textbf{Q}}(\textbf{q}')-\int\frac{d^2\textbf{q}'}{(2\pi)^2}
V_{B-S}(\textbf{q},\textbf{q}';\textbf{Q},E')\widetilde{\chi}_{\textbf{Q}}(\textbf{q}')
=E' \widetilde{\chi}_{\textbf{Q}}(\textbf{q}),
 \label{ZKE}
\end{eqnarray}
In what follows, the last term in (\ref{ZKE}) will be referred as
the B-S term:
\begin{eqnarray}
&V_{B-S}(\textbf{q},\textbf{q}';\textbf{Q},E')=
\frac{[E_v(\textbf{q}'-\alpha_v
\textbf{Q})-E_v(\textbf{q}-\alpha_v
\textbf{Q})]\Omega_c^{\textbf{B}}(\textbf{Q},\textbf{q},\textbf{q}')}{E'-
E_c(\textbf{q}'+\alpha_c \textbf{Q})-E_v(\textbf{q}-\alpha_v
\textbf{Q})} \nonumber\\&+ \frac{[E_c(\textbf{q}'+\alpha_c
\textbf{Q})- E_c(\textbf{q}+\alpha_c
\textbf{Q})]\Omega_v^{\textbf{B}}(\textbf{Q},\textbf{q},\textbf{q}')}{E'-
E_c(\textbf{q}+\alpha_c \textbf{Q})-E_v(\textbf{q}'-\alpha_v
\textbf{Q})}\nonumber\\&+\Omega_{cv}^{\textbf{B}}(\textbf{Q},\textbf{q},\textbf{q}')\left[\frac{1}{E'-
E_c(\textbf{q}'+\alpha_c \textbf{Q})-E_v(\textbf{q}-\alpha_v
\textbf{Q})}+\frac{1}{E'- E_c(\textbf{q}+\alpha_c
\textbf{Q})-E_v(\textbf{q}'-\alpha_v \textbf{Q})}\right].
\label{Veff}
\end{eqnarray}
Here, the following notations have been used:
\begin{equation}\Omega_{c,v}^{\textbf{B}}(\textbf{Q},\textbf{q},\textbf{q}')=\int
d^2\textbf{r}\exp\left[\imath(\textbf{q}'-\textbf{q})\textbf{.r}\right]
\Omega_{c,v}^{\textbf{B}}(\textbf{Q},\textbf{q}';\textbf{r})\label{Omc},\end{equation}
\begin{equation}\Omega_{cv}^{\textbf{B}}(\textbf{Q},\textbf{q},\textbf{q}')=\int
d^2\textbf{r}\exp\left[\imath(\textbf{q}'-\textbf{q})\textbf{.r}\right]
\Omega_{c}^{\textbf{B}}(\textbf{Q},\textbf{q}';\textbf{r})\Omega_{v}^{\textbf{B}}(\textbf{Q},\textbf{q}';\textbf{r})\label{Omcv}.\end{equation}
In position representation, the B-S term generates a
 non-local potential  which depends on the energy $E'$:
\begin{equation}
V_{B-S}(\textbf{r},\textbf{r}';\textbf{Q},E')=
\int\frac{d^2\textbf{q}}{(2\pi)^2}\int\frac{d^2\textbf{q}'}{(2\pi)^2}V_{B-S}(\textbf{q},\textbf{q}';\textbf{Q},E')
\exp[\imath\left(\textbf{q.r}-\textbf{q}'\textbf{.r}'\right)].
\label{VeffR}
\end{equation}
The solution of Eq.
 (\ref{ZKE}) can be written as
$$\widetilde{\chi}_{\textbf{Q}}(\textbf{q})=\exp\left(-\imath
\textbf{q.}\textbf{R}_0\right)\Psi\left(\textbf{q}-\textbf{Q}_0\right),$$
where  the function $\Psi\left(\textbf{q}\right)$ satisfies the
following equation:
\begin{eqnarray}
&E'
\Psi(\textbf{q})=\frac{\hbar^2\textbf{q}^2}{2\mu}\Psi\left(\textbf{q}\right)
-\imath\frac{\gamma \hbar e}{2\mu
c}\left(\textbf{B}_\perp\times\textbf{q}\right)\textbf{.}\nabla_\textbf{q}\Psi\left(\textbf{q}\right)-
\frac{\hbar\Omega}{8R^2}\nabla^2_\textbf{q}\Psi\left(\textbf{q}\right)\nonumber\\&
-\frac{2\pi
e^2}{\epsilon_{\infty}}\int\frac{d^2\textbf{q}'}{(2\pi)^2}\exp\left[\imath
\left(\textbf{q}-\textbf{q}'\right)\textbf{.R}_0 \right]
\frac{f(|\textbf{q}-\textbf{q}'|)}{|\textbf{q}-\textbf{q}'|}\Psi(\textbf{q}')\nonumber\\
&-\int\frac{d^2\textbf{q}'}{(2\pi)^2}\exp\left[\imath
\left(\textbf{q}-\textbf{q}'\right)\textbf{.R}_0 \right]
V_{B-S}(\textbf{q}+\frac{\gamma}{2}\textbf{Q}_0,\textbf{q}'+\frac{\gamma}{2}\textbf{Q}_0;\textbf{Q},E')\Psi(\textbf{q}')
.
 \label{ZKE1}
\end{eqnarray}
 The B-S equation (\ref{ZKE1})
differs from the Schr\"{o}dinger equation. If we neglect the B-S
term in the right-hand side of (\ref{ZKE1}), we  obtain the
Schr\"{o}dinger equation for magnetoexcitons with the Hamiltonian
(\ref{Hperp}). It can be seen that according to the
Schr\"{o}dinger equation, the magnetoexciton dispersion is totally
determined by the Coulomb term, while according to the B-S
equation, the effective potential
(\ref{VeffR}) also contributes to the magnetoexciton dispersion.\\
Since  the Bethe-Salpeter term plays an important role in
determining the magnetoexciton dispersion (see the next two
Sections), one may well ask a question about the physical meaning
of this term. The answer is that the B-S term takes into account
the contributions to the single-particle Green's functions
(\ref{GF1}) from the Landau levels with $n\geq 1$.
\section{Magnetoexciton dispersion in single $GaAs/Al_xGa_{1-x}As$ quantum well}
 \begin{table}
\begin{tabular}{cccccccc}
$L (nm) $ & $B (T) $ & $\beta$ & $E_{c0}(meV)$ & $E_{v0}(meV)$&
$E_{var}(eV)$ & $E_{exp}(eV)$ & $E_{S}(eV)$

\\ \hline
4.03  & 0  & 0.786  & 100& 26.9  & 1.6355 & 1.638 & 1.6355 \\
4.03  & 2 & 0.810  & 100& 26.9 & 1.6356 & 1.639 & 1.6357  \\
4.03  & 4  & 0.869  & 100& 26.9  & 1.6365 & 1.640 & 1.6367  \\
4.32  & 0  & 0.776  & 93.5 & 24.3 & 1.6262 & 1.630 & 1.6262  \\
4.32  & 2  & 0.802  & 93.5 & 24.3  & 1.6265 & 1.631 & 1.6266 \\
4.32  & 4  & 0.861  & 93.5 & 24.3 & 1.6274 & 1.632 & 1.6275 \\
7.2   & 0  & 0.702  & 51.0 & 11.0  & 1.5716 & 1.571 & 1.5716 \\
7.2   & 2  & 0.734  & 51.0 & 11.0  & 1.5719 & 1.572 & 1.5720  \\
7.2   & 4  & 0.803  & 51.0 & 11.0  & 1.5730 & 1.573 & 1.5731 \\
\end{tabular}
\caption{Variational calculations of the heavy-hole exciton
ground-state energies with $\textbf{Q}=0$ for various well widths
$L$ and weak magnetic fields $B$. The trial function (\ref{VF1})
depends on the variational parameter $\beta$. The energy gap is
$E_g=1.519$ eV. The electron and hole confinement energy levels
$E_{c0}$ and $E_{v0}$ are calculated assuming squared-well
potentials of finite depths. The $E_{var}$-column represents the
results from the variational calculations with the following
Luttinger parameters: $\gamma_1=7.36$ and $\gamma_2=2.57$
\cite{LP1}. The measured ground state energies $E_{exp}$ are
reproduced from \cite{Ko}. The $E_{S}$-column represents the
ground-state energies calculated according to the Schr\"{o}dinger
equation with the Hamiltonian (\ref{Hperp})}
\end{table}
In this Section, we first calculate the ground-state energy of a
heavy-hole magnetoexciton with a zero wave vector
($\textbf{Q}=0$), assuming a single GaAs quantum well with a
thickness $L$ sandwiched between two $Al_xGa_{1-x}As$ layers. The
electron in-plane mass $m_c$ and the electron z-mass $m_{cz}$ are
chosen to be $m_c=m_{cz}=0.067 m_0$, where $m_0$ is the bare
electron mass. The in-plane heavy-hole mass $m_v$ and the hole
z-mass $m_{vz}$ are expressed in terms of the Luttinger parameters
$\gamma_1$ and $\gamma_2$: $m_v=m_0/(\gamma_1+\gamma_2)$ and
$m_{vz}=m_0/(\gamma_1-2\gamma_2)$. It is known that the difference
between the bandgap energies of $GaAs$ and $Al_xGa_{1-x}As$
provides a finite potential well, confining the electron-hole
pairs in the Galas quantum well. We assume that the potentials are
square-well potentials of finite depths $V_c=0.6\Delta Eg(x)$ and
$V_v=0.4\Delta Eg(x)$, respectively. The energy-band-gap
discontinuity\cite{LP1} is assumed to be $\Delta
Eg(x)=(1.555x+0.37x^2) meV$. The confinement energy levels
$E_{c0}$ and $E_{v0}$ are obtained by solving the following
transcendental equations:
$$\tan\left(\frac{L}{2a_B}\sqrt{\frac{m_{cz} E_{c0}}{\mu
E_B}}\right)= \sqrt{\frac{V_c} {E_{c0}}-1},$$
$$\tan\left(\frac{L}{2a_B}\sqrt{\frac{m_{vz} E_{v0}}{\mu
E_B}}\right)=\sqrt{\frac{V_v}{E_{v0}}-1}.$$ Here,
 $E_B=\hbar^2/2\mu
a^2_B$ is the exciton Bohr energy. The structure factor $f(k)$ is
calculated by means of the following wave functions:
\begin{eqnarray}
&\psi^0_{c,v}(z)=A_{c,v}\exp\left[z\frac{L}{a_B}\sqrt{\frac{m_{cz,vz}(V_{c,v}-E_{c0,v0})}
{\mu E_B}}\right],\quad -\infty<z<-1/2,\nonumber\\&
\psi^0_{c,v}(z)=B_{c,v}\cos\left(z\frac{L}{a_B}\sqrt{\frac{m_{cz,vz}E_{c0,v0})}{\mu
E_B}}\right),\quad -1/2<z<1/2,\nonumber\\&
\psi^0_{c,v}(z)=A_{c,v}\exp\left[-z\frac{L}{a_B}\sqrt{\frac{m_{cz,vz}(V_{c,v}-E_{c0,v0})}{\mu
E_B}}\right],\quad 1/2<z<\infty,\nonumber\\&
B_{c,v}=\left[\frac{1}{2}+a_B/\left(L\sqrt{\frac{m_{cz,vz}(V_{c,v}-E_{c0,v0})}{\mu
E_B}}\right)\right]^{-1/2},\nonumber\\&
 A_{c,v}=B_{c,v}\exp\left[\frac{L}{2a_B}\sqrt{\frac{m_{cz,vz}(V_{c,v}-E_{c0,v0})}{\mu
E_B}}\right]\cos\left(\frac{L}{2a_B}\sqrt{\frac{m_{cz,vz}E_{c0,v0})}{\mu
E_B}}\right).\nonumber\end{eqnarray} Since the B-S equation
(\ref{ZKE1}) is rather complicated we shall obtain
 numerical results for the ground-state energy  within the framework of the
variational approach. In the case of weak magnetic fields, i.e.
$\hbar\Omega<<E_B$, we use a hydrogen-like trial function with a
variational parameter $\beta$:
\begin{equation}\psi_\beta(r)=\frac{2\sqrt{2}\beta}{\sqrt{\pi}a_B}\exp\left(-\frac{2r\beta}
{a_B}\right).\label{VF1}\end{equation}
 With this trial function we calculate the following magnetoexciton energy:
 $$E=E_g+E_{c0}+E_{v0}-E(\beta)E_B,$$
 where $E(\beta)$ is defined by the solution of the following equation:
 \begin{equation}
E(\beta)=-4\beta^2+128\beta^3\int_{0}^{\infty}dx \frac{f(x\frac{L}{a_B})}{(16\beta^2+x^2)^{3/2}}\\
-\frac{3}{128\beta^2}\left(\frac{\hbar\Omega}{E_B}\right)^2+V_{B-S}(\beta,E,B),
\label{LMF} \end{equation} With the trial function (\ref{VF1}),
the B-S contribution to the ground
 state is:
 \begin{eqnarray}
 &V_{B-S}(\beta,E,B)=\frac{\hbar\Omega}{E_B}\quad\frac{a_B^2(1-\gamma^2)}{2^{12}E^2\beta^4(a_B^2E-2\beta^2)^7} \{(a_B^2E-2\beta^2)[15a_B^{14}E^7
 -162a_B^{12}E^6\beta^2\nonumber\\
 &+8a_B^8E^4\beta^6(-195+896E^2-36\gamma^2)-4a_B^{10}E^5\beta^4(-173+128E^2+4\gamma^2)\nonumber\\&+64a_B^4E^2\beta^{10}(41+1408E^2-322\gamma^2-492\gamma^4)\nonumber\\
&-512a_B^2E\beta^{12}(3+208E^2-18\gamma^2+15\gamma^4)-32a_B^6E^3\beta^8(79+1152E^2+802\gamma^2\nonumber\\&+172\gamma^4)+1024\beta^{14}[48E^2+(-1+\gamma^2)^2]]\nonumber\\
&-64E^2\beta^8[-2048a_B^2E\beta^6+1024\beta^8+48a_B^4\beta^4(1+32E^2+\gamma^2-12\gamma^4)\nonumber\\
&-16a_B^6E\beta^2(3+32E^2+24\gamma^2(2+\gamma^2)\nonumber\\
 &+a_B^8E^2(64E^2-3[11+8\gamma^2(7+\gamma^2)])]\ln\left(\frac{a_B^2E}{2\beta^2}\right)\}.
 \label{VeffL}\end{eqnarray}The dimensionless variables $E$ and $a_B$ in the right-hand side of Eq.(\ref{VeffL})
 must be replaced by $E(\beta)\hbar\Omega/E_B^2$ and $a_B/l$,
 respectively.
The results obtained by using the hydrogen-like trial function are
presented in Table 1. We used more significant figures to stress
on the fact that
 the magnetoexciton energies, calculated by applying the B-S formalism are
 extremely closed to those, provided by the Schr\"{o}dinger
 equation.\\
 The magnetoexciton dispersion are determined by the Coulomb interaction and the B-S term in Eq. (\ref{ZKE1}).
The contribution from the Coulomb interaction to the energy of the
magnetoexciton (in $E_B$ units) increases quadratically for small
wave vectors $Qa_B<<1$, and can be written as $(Qa_B)^2\mu /M_C$.
The hydrogen-like trial function provides the following expression
for the in-plane exciton mass $M_C$:
$$\frac{\mu}{M_C}=32\beta^3\left(\frac{R}{a_B}\right)^4\int_{0}^{\infty}dx \frac{x^2f(x\frac{L}{a_B})}{(16\beta^2+x^2)^{3/2}}.$$
\begin{table}
\begin{tabular}{ccccccccccccc}
$L (nm) $ & $B (T) $ & $\beta $ & $E_{var}(eV)$ & $E_{exp}(eV)$ &
$E_{S}(eV)$ & $M_{C}/m_0$ & $M_{B-S}/m_0$

\\ \hline
4.03  & 20  & 0.85  &  1.650 & 1.644 & 1.651 &  0.145 & 0.0025\\
4.03  & 18  & 0.84   &  1.648 & 1.643 & 1.649&  0.127 & 0.0010\\
4.03  & 16  & 0.84  &  1.647 & 1.642 & 1.647 &  0.114 & 0.0002\\
4.32  & 20  & 0.84   &  1.641 & 1.636 & 1.642&  0.147 & 0.0026\\
4.32  & 18  & 0.83   &  1.639 & 1.635 & 1.640&  0.129 & 0.0011\\
4.32  & 16  & 0.83  &  1.638 & 1.634 & 1.638&  0.116 & 0.0002\\
7.2   & 20  & 0.86   &  1.587 & 1.583 & 1.588&  0.176 & 0.0044\\
7.2   & 18  & 0.84   &  1.585 & 1.582 & 1.586&  0.159 & 0.0022\\
7.2   & 16  & 0.84   &  1.583 & 1.581 & 1.584&  0.142 & 0.0007\\
7.49  & 20  & 0.86   &  1.584 & 1.580 & 1.584&  0.178 & 0.0046\\
7.49  & 18  & 0.84  &  1.582 & 1.579 & 1.582&  0.161 & 0.0024\\
7.49  & 16  & 0.84  &  1.580 & 1.578 & 1.580&  0.144 & 0.0008\\
7.5   & 14.5  & 0.67  &  1.577 & 1.577 & 1.572&  0.131 & 0.0302\\
7.5   & 12  & 0.64  &  1.575 & 1.573 & 1.570&  0.049 & 0.0160\\
7.5   & 8.5  & 0.60   &  1.572 & 1.570 & 1.569&  0.026 & 0.0071\\
\end{tabular}
\caption{Variational calculations of the heavy-hole exciton
ground-state energies for various well widths $L$ and strong
magnetic fields $B$. The trial function (\ref{VF}) depends on the
variational parameter $\beta$. The energy gap is $E_g=1.519$ eV
for the $L=4.03, 4.32, 7.2,$ and $7.49$-nm wells, and $E_g=1.512$
eV for the $L=7.5$-nm. The $E_{var}$-column represents the
energies obtained by the variational method using the following
Luttinger parameters: $\gamma_1=6.9$ and $\gamma_2=2.4$ \cite{LP}.
The measured ground state energies $E_{exp}$ for the $L=4.03,
4.32, 7.2,$ and $7.49$-nm wells are reproduced from \cite{Ko}, and
for the $L=7.5$-nm well from \cite{Ro}. The $E_{S}$-column
represents the ground-state energies calculated according to the
Schr\"{o}dinger equation. The $M_{C}$ and $M_{B-S}$
 are the masses calculated according to Eqs. (\ref{MasssC}) and
(\ref{BSterm2}).}
\end{table}  The contribution to the exciton dispersion due to the B-S
  term can be evaluated analytically. We found that it also increases quadratically for small
wave vectors, but for $B<4T$, this contribution is about one tenth
of $(Qa_B)^2\mu /M_C$. Thus, in a weak magnetic
  field, there is no measurable difference between the results
  calculated by the Schr\"{o}dinger
  equation, and these obtained by the more complicated B-S formalism. For a weak perpendicular magnetic field and small wave vectors, the
  Coulomb interaction dominates, which means that a hydrogen type of
  ground state slightly modified by the magnetic field exists.
\\Next, we consider the case of a strong magnetic field. In
this regime we choose the trial wave function $\psi_\beta(r)$ to
be similar to the corresponding ground-state wave function of a
charge particle in a magnetic field, but depending on a
variational parameter $\beta$:
\begin{equation}\psi_\beta(r)=\frac{1}{\sqrt{2\pi}\beta}\exp\left(-\frac{r^2}
{4\beta^2}\right).\label{VF}\end{equation} Here, and in what
follows, we use the exciton cyclotron energy $\hbar\Omega$
 for energy unit and magnetic length $R$ for unit
 length. The ground state magnetoexciton energy will be calculated by
minimizing the energy functional
$E'(\beta)=(E-E_g-E_{0c}-E_{0v})/\hbar\Omega$ with respect to the
variational parameter $\beta$:
\begin{equation}E'=\frac{1}{4}\left(\frac{1}{\beta^2}+\beta^2\right)+V_C(\beta)+V_{B-S}(\beta,E')+V_C(\beta,\textbf{Q})
+V_{B-S}(\beta,E',\textbf{Q}).\label{EnFS}\end{equation}
 Note, that (i) all terms in the last equation are dimensionless (in a cyclotron energy $\hbar\Omega$ unit), and (ii)
  we have written the contributions from the Coulomb interaction and from the
 B-S term (\ref{Veff}) as a sum of $Q$-independent terms, $V_C(\beta)$ and $V_{B-S}(\beta,E')$, and
  $Q$-dependent terms, $V_C(\beta,\textbf{Q})$ and $V_{B-S}(\beta,E',\textbf{Q})$. The $Q$-dependent terms will be used to obtain the
  magnetoexciton dispersion. The second and the third term in (\ref{EnFS}) are given by:
\begin{equation}V_C(\beta)=-\frac{E_b}{\hbar\Omega }\sqrt{\frac{2}{\pi}}\int_0^\infty dxf(x\frac{L}{R})
\exp(-\frac{x^2\beta^2}{2}),\label{VC}\end{equation}
\begin{eqnarray}
&V_{B-S}(\beta,E')=\frac{e^{-4E'\beta^2}\beta^2(-1+\gamma^2)}{64E'^2}
\{e^{4E'\beta^2}[-56E'^2\beta^4\gamma^4+32E'^3\beta^6\gamma^4+(-1+\gamma^2)^2\nonumber\\&+4E'\beta^2(-1-2\gamma^2+3\gamma^4)]\nonumber\\
&-32E'^2\left[-1+\beta^4\gamma^2\left[-1+(3+4E'\beta^2(-2+E'\beta^2))\gamma^2\right]\right]\texttt{Ei}(4E'\beta^2)\}.\label{VeffQ0}
\end{eqnarray}
Here, $E_b=\sqrt{\pi/2}e^2/(\epsilon_\infty R)$ is the binding
energy of the two-dimensional ($L=0,\beta=1$) magnetoexciton,
calculated according to the Schr\"{o}dinger equation.\\The energy
of the magnetoexciton increases quadratically for small wave
vectors ($QR<<1$):
$V_C(\beta,\textbf{Q})=[\mu/2M_C(L,B,\beta)](QR)^2$ and
$V_{B-S}(\beta,E',\textbf{Q})=[\mu/2M_{B-S}(L,B,\beta)](QR)^2$.
The in-plane mass $M_C(L,B,\beta)$ is due to the Coulomb
interaction and does not depend on the electron or the hole mass:
\begin{equation}
\frac{M_{2D}}{M_C(L,B,\beta)}= \sqrt{\frac{2}{\pi}}\int_0^\infty
dxf(x\frac{L}{R})x^2
\exp(-\frac{x^2\beta^2}{2}),\label{MasssC}\end{equation} where
$M_{2D}=2^{3/2}\epsilon_\infty \hbar^2/(\sqrt{\pi}e^2R)$. The
second in-plane mass, $M_{B-S}$, has its origin in the fact that
the B-S term depends on $\textbf{Q}$, and for $QR<<1$, $M_{B-S}$
is defined by the following equation:
\begin{eqnarray}&
\frac{\mu}{2M_{B-S}(L,B,\beta)}=
\frac{e^{-4E'\beta^2}(-1+\gamma^2)}{256E'^3}\{e^{4E'\beta^2}[256E'^5\beta^{12}\gamma^6+64E'^4\beta^{10}\gamma^4(5-17\gamma^2)\nonumber\\&
-3\beta^2(-1+\gamma^2)^3-2E'\beta^4(-1+\gamma^2)^2(1+12\gamma^2)-48E'^2\beta^6\gamma^2(2-7\gamma^2+5\gamma^4)\nonumber\\&+16E'^3
[-2+2\beta^4+\beta^8\gamma^2(4-53\gamma^2+74\gamma^4)]]\nonumber\\&-64E'^3\beta^2[-\beta^2+16E'^3\beta^{12}\gamma^6
+4E'^2\beta^{10}\gamma^4(5-18\gamma^2)+\beta^6\gamma^2(-7+33\gamma^2-30\gamma^4)\nonumber\\&
+2E'\left[-1+\beta^4+\beta^8\gamma^2(2-29\gamma^2+45\gamma^4)\right]]\texttt{Ei}(4E'\beta^2)
\},\label{BSterm2}
\end{eqnarray}
where $\texttt{Ei}(x)=-\int_{-x}^\infty dt\exp(-t)/t$ is the
exponential integral function (the principle value of the
integral is taken).\\
 Table 2 gives the results of our
variational calculations. It can be seen that the B-S equation
provides similar results for the ground-state energies as the
Schrodinger equation does. Since the B-S mass is much smaller than
the Coulomb mass, one can say that in strong magnetic fields the
exciton dispersion for small wave vectors ($QR<<1$) is determined
by the B-S term rather than the Coulomb interaction.
\section{Coupled quantum wells in strong magnetic fields}
In this Section, we consider exactly the same double well
electron-hole system as in Refs. \cite{LZ,LZ1}. The electron layer
and hole layer have finite widths, denoted below by
$\textrm{L}_{c}$ and $\textrm{L}_{v}$, and they are separated by a
distance $\textrm{D}$. We assume that the electrons and holes are
confined between two parallel, infinitely high potential barriers.
This assumption greatly simplifies our numerical calculations of
the magnetoexciton energy and the Coulomb mass, but by neglecting
the existence of the finite confinement potentials, we cannot
provide a more realistic value for this part of the exciton energy
related to the exciton confinement along z-direction, than the sum
of the well-known terms $\hbar^2\pi^2/2m_{c,v}L^2_{c,v}$.
Obviously, the more realistic model of a symmetric (or asymmetric)
DQW with finite quantum-well widths\cite{B,B1} will cause minor
corrections to our main conclusions, which are: (1) the B-S
formalism provides a term, which does not exists in the
Schrodinger equation, and (2) the term plays an important
role in determining the magnetoexciton dispersion.\\
The basic features of the CQW's magnetoexcitons are the same as
that of the SQW magnetoexcitons. However, because of the
separation between the electron and hole layers, the Coulomb
energy and the Coulomb in-plane mass differ quantitatively from
those of the SQW magnetoexciton. In other words, in strong
magnetic fields, Eq. (\ref{EnFS}) holds, but the Coulomb
interaction and the corresponding in-plane mass are defined as
follows:
\begin{eqnarray}&V_C(\beta)=-\frac{E_b}{\hbar\Omega }\sqrt{\frac{2}{\pi}}\int_0^\infty
 dxe^{-\frac{x^2\beta^2}{2}}F\left(x,
 \frac{L_c}{R},\frac{L_v}{R},\frac{D}{R}\right),\\&
 \frac{M_{2D}}{M_C(L,B,\beta)}=
\sqrt{\frac{2}{\pi}}\int_0^\infty
dxx^2e^{-\frac{x^2\beta^2}{2}}F\left(x,
\frac{L_c}{R},\frac{L_v}{R},\frac{D}{R}\right).\label{CQWMass}\end{eqnarray}
In CQW's, the structure factor is:
$$F(x,\xi_c,\xi_v,d)=\frac{16\pi^4(1-e^{-\xi_cx})(1-e^{-\xi_vx})e^{-dx}}{\xi_c\xi_vx^2(4\pi^2+\xi_c^2x^2)(4\pi^2+\xi_v^2x^2)}.$$
Table 3 gives the result of our numerical calculation of the
magnetoexciton energy, but relatively to the  $E_g+E_{0c}+E_{0v}$
level. We used the same parameters as in Refs. \cite{LZ} and
\cite{RG}. It can be seen that the B-S equation provides slightly
different results for the binding
energy than the Schr\"{o}dingier equation.\\
\begin{table}
\begin{tabular}{ccccccc}
$B (T) $ &$\beta$ & $E_{var}(meV)$ & $E_{S}(meV)$ &
$\frac{M_C}{m_0}$ & $\frac{M_{B-S}}{m_0}$

\\ \hline
10 &0.96 & 6.36  & 6.56  & 2.06 &0.228\\
9  &0.96 & 5.17  & 5.43  &  1.75 &0.221\\
8  &0.96 & 4.03  & 4.31  & 1.46 & 0.216\\
7  &0.96 & 2.94  & 3.20  &1.19 &0.215\\
6  &0.95 & 1.91  & 2.11  &0.95& 0.218\\
5  &0.94 & 0.95  & 1.04  &0.72 &0.230\\
4  &0.92 & 0.01  & 0.01  &0.52 &0.247\\
\end{tabular}
\caption{Variational calculations of the magnetoexciton energies
for various strong magnetic fields $B$, measured relatively to the
 $E_g+E_{0c}+E_{0v}$ level. The trial function (\ref{VF}) depends
on the variational parameter $\beta$. The $E_{var}$-column
contains the energies calculated by the variational method with
the following parameters: $m_c=0.067 m_0$, $m_v=0.18 m_0$,
$\epsilon_\infty=12.35$, $L_c=L_v= 8nm$, $D=11.5 nm$. The
$E_{S}$-column represents the magnetoexciton energies calculated
according to the Schr\"{o}dinger  equation. $M_C$ is the in-plane
mass defined by Eq. (\ref{CQWMass}). The $M_{B-S}$ is the mass
calculated according to Eq. (\ref{BSterm2}).}
\end{table}
The main difference between the B-S and the Schr\"{o}dinger
equation is in their predictions about the in-plane magnetoexciton
mass in a strong magnetic field. Unfortunately, optical
experimental studies can provide information about the exciton
dispersion only for $ Q\leq Q_{ph}$, where $\hbar Q_{ph}$ is the
photon momentum. Other studies, such as the photoluminescence
measurement experiments which can measure the exciton-mass
dependence of the recombination time, or experimental data related
to the polariton effects, can provide information about the
magnetoexciton dispersion. Many of these experimental
techniques\cite{InPl,InP2,InP3,InP4,InP5} are used to measure the
magnetoexciton dispersion in the presence of an in-plane magnetic
field. As we mentioned above, the measurable differences between
the magnetoexciton dispersions, as predicted by the B-S formalism
and by the Schr\"{o}dinger  equation, are to be expected in strong
perpendicular magnetic fields. To the best of our knowledge, there
is only one paper\cite{LZ} where the exciton dispersion in
$GaAs/Ga_{0.67}Al_{0.33}As$ CQW's in a weak perpendicular magnetic
field has been measured. There is a good agreement between the
mass $M_C$ and the measured mass in a weak magnetic field.
Referring to the conclusion that the B-S term in a weak magnetic
field has a very small contribution to the dispersion compare to
the contribution due to the Coulomb interaction, one can say that
there exists
a good agreement between the B-S formalism and the measurements.\\
Next, we discuss the fact that $M_C$ increases by about 4 times if
we increase the magnetic field from 4T to 10T.  If the
magnetoexciton dispersion in strong magnetic fields ($B>5T$) is
determined mainly by the B-S term, then the magnetoexciton mass
should not increase so dramatically, and therefore, new
experimental points are needed to prove or disprove the
conclusions drawn by applying the B-S formalism.
\section{Conclusion}
We have applied the B-S formalism to the quantum-well excitons in
a constant magnetic field applied along the axis of growth of the
quantum-well structure. We found that (1) in the LLL approximation
the B-S equations provides the same results as the Schr\"{o}dinger
equation; (2) beyond the LLL approximation, the B-S equation
contains an extra term (B-S term). This term takes into account
the transitions to the Landau levels with indexes $n\geq 1$. We
applied a variational procedure to obtain the effect of the B-S
term on the magnetoexciton ground-state energy and magnetoexciton
mass. We used a simple hydrogen-like trial wave function in a weak
magnetic field, and figured out that in a weak perpendicular
magnetic field the results obtained by the B-S formalism are very
close to the results calculated by means of the Schr\"{o}dinger
equation. In a strong magnetic field, we used a trial function
similar to the wave function of a charged particle in a magnetic
field. We calculated that in a strong magnetic field, the
ground-state energy is very close to that obtained by means of the
Schr\"{o}dinger  equation, but  the magnetoexciton dispersion is
determined by the B-S term rather than the electron-hole Coulomb
term in the Schr\"{o}dinger equation.


\begin{thebibliography}{99}
\bibitem{Shevchenko97}
S.~I.Shevchenko, \emph{Phys. Rev. B} \textbf{56}, 10355 (1997).

  \bibitem{Lozovik97}
Yu. Lozovik, and A.~M. Ruvisky, \emph{Zh. Eksp. Teor. Fiz.}
\textbf{112}, 1791 (1997) [\emph{Sov.Phys. JETP} \textbf{85}, 979
(1997)].

\bibitem{Dzyubenko}
A.~B. Dzyubenko, \emph{JETP Lett.} \textbf{66},
 617 (1997).

\bibitem{Shevchenko98}
S.I. Shevchenko, \emph{Phys. Rev. B} \textbf{57},
  14809 (1998).


\bibitem{Lozovik99}
  Yu. E.  Lozovik, O.~L. Berman, and V.~G. Tsvetus, \emph{Phys. Rev. B} \textbf{59}, 5627 (1999).


\bibitem{BS}  E. E. Salpeter and H. A. Bethe, \emph{Phys. Rev.}
{\bf 84}, 1232 (1951).
%
\bibitem{GL}M. Gell-Mann and F. Low, \emph{Phys. Rev.}
{\bf 84}, 350 (1951).
%
\bibitem{W} C. G. Wick, \emph{Phys. Rev.}
{\bf 96}, 1124 (1954).
%
\bibitem{C}
 R. E. Cutkosky, \emph{Phys. Rev.} {\bf 96}, 1135
(1954).
%
\bibitem{P}E. A. Manykin, M. I. Ozhovan, and P. P. Poluektov,  \emph{Teor. Mat. Fiz.}
\textbf{49}, 283 (1981) [\emph{Theor. Math. Phys.} \textbf{49},
1035 (1981)].
%
\bibitem{MC} V. P. Gusynin, V. A. Miransky, and I. A. Shovkovy, \emph{Phys. Rev.
Lett.} \textbf{73}, 3499 (1994).
%
\bibitem{MC1} V. P. Gusynin, V. A. Miransky, and I. A. Shovkovy, \emph{Phys. Rev.
D} \textbf{52}, 4718 (1995).
%
 \bibitem{L} C. N. Leung, Y. J. Ng, and A. W. Ackley, \emph{Phys. Rev. D}
\textbf{54}, 4181 (1996).
%
 \bibitem{H} D. K. Hong, Y. Kim and S.-J. Sin, \emph{Phys.
Rev. D} \textbf{54}, 7879 (1996).
%
\bibitem{FM} K. Farakos and N. E. Mavromatos, \emph{Int. J. Mod. Phys. B}
\textbf{12}, 809 (1998).

%
\bibitem{DL} D. Lehmann, \emph{Communications in Math. Phys.} {\bf 173}, 155 (1995).
%
\bibitem{HA} R. London, \emph{J. Phys.} \textbf{27}, 649 (1960).
%
\bibitem{HA1} R. J. Elliott, and R. London, \emph{J. Phys. Chem.
Sol.} \textbf{15}, 196 (1960).

\bibitem{HA2} H. Hasegawa, and R.E. Howard, \emph{J. Phys. Chem.
Sol.} \textbf{21}, 179 (1961).
%
\bibitem{HA3}L. P. Gor'kov, and I. E. Dzyaloshinskii, \emph{Zh. Eksp. Teor. Fiz.}
\textbf{53}, 717 (1967)[\emph{Sov. Phys. JETP} \textbf{26}, 449
(1968).
%
\bibitem{Z1} Z. G. Koinov, \emph{Phys. Rev. B} \textbf{65}, 155332
(2002).
%
\bibitem{Z2} Z. Koinov, \emph{Phys. Rev. B} \textbf{77}, 165333
(2008).
%
\bibitem{LP1} R.L. Greene, K.K. Bajaj and D.E. Phelps, \emph{Phys. Rev. B} \textbf{29},  1807 (1984).
%
\bibitem{Ko} H. S. Ko, S. J. Rhee, Y. M. Kim, W. S. Kim, D. H. Kim, J. H. Bae, Y. S. Kim, J. C. Woo,
D. W. Kim, and T. Schmiedel, \emph{Superlattices and
Microstructures}  {\bf 24},  259  (1998).
%
%
\bibitem{LP} N. Peyghambarian, S.W. Koch and A. Mysyrowicz, Introduction to
Semiconductor Optics (Prentice Hall, New Jersey, 1993) p.173.
%

\bibitem{Ro} D. C. Rogers, J. Singleton, R. J. Nicholas, C. T. Foxon, and K. Woodbridge,
Phys. Rev. B,  {\bf 34},  4002  (1986).
%
\bibitem{LZ} L.V. Butov, C.W. Lai, D.S. Chemla, Y. E. Lozovik, K.L. Campman, and A.C. Gossard,
\emph{Phys. Rev. Lett. }{\bf 87}, 216804 (2001).
%
\bibitem{LZ1} Yu.E. Lozovik,
I.V. Ovchinnikov, S. Yu. Volkov, L.V. Butov, and D.S. Chemla,
\emph{Phys. Rev. B }{\bf 65}, 235304 (2002).
%
\bibitem{B} F. Vera and Z. Barticevic, \emph{Journal of Appl. Phys.}
\textbf{83}, 7720 (1998).
%
\bibitem{B1} B. Flores-Desirena and F. P\'{e}rez-Rodríguez, \emph{Phys.
Stat. Sol. (c)} \textbf{1}, S38 (2004).
%
\bibitem{RG} E. Reyes-Gomez, L. E. Oliveira, and M. de Dios-Leyva, \emph{Phys. Rev.
B} {\bf 71}, 045316 (2005).
%
\bibitem{InPl} D. M. Whittaker, T. A. Fisher, P. E. Simmonds, M. S. Skolnick, and  R. S. Smith, \emph{Phys. Rev.
Lett.} \textbf{67},  887 (1991).
\bibitem{InP2} L. V. Butov, A. V.
 Mintsev, Y. E. Lozovik, K. L. Campman, and A. C. Gossard, \emph{Phys. Rev.
 B}
\textbf{62}, 1548 (2000).
\bibitem{InP3} A. Parlangeli, P. C. M. Christianen, J.
C. Maan, I. V. Tokatly, C. B. Soerensen, and P. E. Lindelof,
\emph{ Phys. Rev. B} \textbf{62}, 15323 (2000).
\bibitem{InP4} M. Orlita
\emph{et al.}, \emph{Phys. Rev. B }\textbf{70}, 075309 (2004).
\bibitem{InP5} B.
M. Ashkinadze, E. Linder, E. Cohen, and L. N. Pfeiffer,
\emph{Phys. Rev. B} \textbf{71}, 045303 (2005).
%

\end{thebibliography}
\end{document}